\documentclass[aps,prb,twocolumn,groupedaddress,showpacs]{revtex4}

\usepackage[bookmarks]{hyperref}
\usepackage[dvips]{graphicx}
\usepackage{amsmath}
\usepackage{amstext}
\usepackage{graphics}
\usepackage{exscale}
\usepackage{epsfig}
\usepackage[T1]{fontenc}
\usepackage{bm}
\usepackage{color}

\usepackage{version}

\usepackage{latexsym}

\usepackage[tight]{subfigure}

\renewcommand{\epsilon}{\varepsilon}

\newcommand{\integral}[3]{\!\int\limits_{#2}^{#3}\!\!{\rm d}#1\;}

\newcommand{\expval}[2]{ \langle  #1 #2\ \!\! \rangle}

\newcommand{\elcre}[2]{ c^{\dagger}_{#1,#2}}
\newcommand{\elann}[2]{ c_{#1,#2}}

\newcommand{\e}{\mathrm e}

\newcommand{\vk}{{\bm k}}
\newcommand{\vq}{{\bm q}}

\newcommand{\Imag}{\mathrm{Im}}

\newcommand{\hc}{\mathrm{h.c.}}

\includeversion{commentst1} 

\begin{document}

\title{Quantitative reliability of Migdal-Eliashberg theory for strong electron-phonon coupling}  
\author{Johannes Bauer${}^1$, Jong E. Han${}^{1,2}$, and Olle
Gunnarsson${}^1$}
\affiliation{${}^1$Max-Planck Institute for Solid State Research,
Heisenbergstr.1,
 70569 Stuttgart, Germany} 
\affiliation{${}^2$  Department of Physics, SUNY at Buffalo, Buffalo,
New York 14260, USA}
\date{\today} 
\begin{abstract}
We reassess the validity of Migdal-Eliashberg (ME) theory for coupled
electron-phonon systems for large couplings $\lambda$. Although 
model calculations have found that ME theory breaks down for $\lambda\sim 0.5$,
it is routinely applied for $\lambda>1$ to strong coupling superconductors.  
To resolve this discrepancy it is important to distinguish between {\em bare} parameters, used as
input in models, and {\em effective} parameters, derived from
experiments. We show explicitly that ME gives accurate results for the
critical temperature and the spectral gap for large effective $\lambda$. This
provides quantitative theoretical support for the applicability of ME theory
to strong coupling conventional superconductors. 
\end{abstract}

\maketitle
\section{Introduction}
The theory of conventional superconductivity, where pairing is mediated
by the coupling of electrons to lattice vibrations, is considered as one of
the major achievements of the twentieth century condensed matter physics. It
is based on the relatively simple Eliashberg equations,\cite{Eli60} where
vertex corrections are neglected.
The application of these equations is justified by Migdal's theorem,\cite{Mig58} which states that vertex corrections are   
proportional to the effective coupling strength $\lambda$ and the ratio
of phonon
$\omega_{\rm ph}$ and electronic energy scale $W$, which usually is of the
order $1/100$
and less.\cite{Car90,MC08} The properties of the superconducting state
in Migdal-Eliashberg (ME) theory are then largely  determined by $\lambda$,
$\omega_{\rm ph}$ and a phenomenological parameter for the Coulomb repulsion
$\mu^*$.  ME theory is also routinely used as a standard pairing theory for
other situations where a bosonic pairing mechanism is analyzed.\cite{CPS08,Dea09}

A number of studies mainly based on the Holstein model, which go beyond the
ME theory, have illustrated that vertex corrections cannot be neglected for
large coupling strength even if the ratio $\omega_{\rm ph}/W$ 
is very small and that the ME theory becomes inaccurate when\cite{differentlambda}
$\lambda$ exceeds a certain
value.\cite{BZ98,Ale01,MHB02,CC03,HD08} 
In the adiabatic limit, Benedetti and Zeyher
\cite{BZ98} found a breakdown of
Migdal's theorem due to the appearance of additional extremal paths in
the action for $\lambda\gtrsim 0.4$.\cite{differentlambda}
Capone and Ciuchi \cite{CC03} found
quantitative deviations of self-consistent ME calculations from DMFT 
already for intermediate coupling strengths and qualitatively different
behavior for stronger coupling.  Alexandrov \cite{Ale01} argued that
even in the
adiabatic limit ME theory breaks down due to bipolaron formation and
symmetry
breaking when $\lambda$ exceeds one. 
For strong coupling superconductors values of $\lambda$ of
the order 1-3 are commonly quoted.\cite{Car90,MC08}
The model calculations therefore suggest that strong coupling superconductors
do not lie within the range of applicability of the ME theory.

The purpose of this paper is to bring the results from the model
studies in a form that they can be compared in a meaningful way to the standard
diagrammatic approach for superconductivity. This allows us to clarify the
quantitative reliability
of ME theory for relevant values of $\lambda$ and $\omega_{\rm ph}/W$. 
If phonon renormalization occurs it is necessary to distinguish the {\it bare} model 
parameters from the {\it effective} parameters describing the phonon properties.\cite{MK82,Mar90,DAM08} 
The latter correspond to the ones derived from experiment or density
functional calculations. We study the Holstein model in the limit where the
lattice has infinite dimensions.  In this limit the dynamical mean field theory
(DMFT) \cite{GKKR96} becomes exact. These DMFT results serve as a benchmark
for ME calculations. 

We show in qualitative agreement with earlier work in the normal phase
\cite{CC03} that self-consistent ME calculations become inaccurate already at
moderate {\it bare} coupling both for electronic and phonon properties.   
However in contrast with previous interpretations, we show that at these bare
couplings the effective coupling is very large, larger than for strong
coupling superconductors.
For effective couplings relevant for strong coupling
superconductors the ME theory is still accurate.  

In many applications of ME theory the phonons are
not calculated self-consistently, but taken as an input either from a
different calculation or experiment. Then one is interested in how accurately
electronic properties are described by the ME equations for a given phonon
spectrum.  
We can  check this explicitely by taking DMFT as a benchmark for electronic
properties and providing the full phonon spectrum as an input for
the ME calculations (termed ME+ph later). 
We show that the electronic properties are predicted very
reliably up to large {\it effective} coupling strengths within ME+ph
calculations, i.e. with an accuracy of better than 10\%.
In this paper we will neglect the effect of the Coulomb
interaction usually taken into account via the parameter $\mu^*$. 

The paper is structured as follows: In Sec. II, we first recall the usual
definition of the pairing function $\alpha^2 F(\omega)$ and the coupling strength $\lambda$. Then
we give explicit details for the DMFT and ME approaches. In Sec. III, results
for the comparison of the DMFT and ME calculations are shown, followed by the
conclusions in Sec. IV. 

\section{Model and formalism}
\subsection{Pairing function}
The pairing function $\alpha^2 F(\omega)$ is an essential ingredient for
conventional superconductivity. It can be defined by
\cite{Car90,MC08} 
\begin{equation}
 \alpha_{\vk,\vk'}^2F(\omega)=\rho_0|g_{\vk,\vk'}|^2\rho^D_{\vk-\vk'}(\omega),
\end{equation}
where $\rho_0$ is the electronic density of states at the Fermi level,
$g_{\vk,\vk'}$ the electron-phonon coupling matrix element and
$\rho^D_{\vq}(\omega)$ the phonon spectral function, related to the
phonon propagator as 
\begin{equation}
 D_{\vq}(i\omega_m)=\integral{\omega}{0}{\infty}\rho^D_{\vq}(\omega)\frac{2\omega}{(i\omega_m)^2-\omega^2}.
\label{phonspec}
\end{equation}
These are the dressed phonon quantities of the interacting system. In
conventional theory these are often taken from experiment or estimated
by a different method, and  then inserted in the Eliashberg equations to
solve
for $T_c$, the spectral gap and other properties. As the properties of
conventional
superconductivity are mostly confined to a small window around the Fermi
energy, often a Fermi surface average is used,
\begin{equation}
 \alpha^2F(\omega)=\frac1{\rho_0^2}\sum_{\vk,\vk'}
\alpha_{\vk,\vk'}^2F(\omega)
\delta(\epsilon_{\vk}-\mu)\delta(\epsilon_{\vk'}-\mu).
\end{equation}
Then the superconducting state is largely determined through the
coupling constant\cite{McM68,AD75,MC08} $\lambda$,
\begin{equation}
 \lambda=2\integral{\omega}{0}{\infty}\frac{\alpha^2F(\omega)}{\omega}.
\label{eq:deflambda}
\end{equation}

\subsection{Holstein model and DMFT approach}
Our quantitative test of the ME theory is based on a model, which has
been
frequently used in the literature, the Holstein model, 
\begin{eqnarray}
 \label{holham}
 H&=&-\sum_{i,j,{\sigma}}(t_{ij}\elcre i{\sigma}\elann
j{\sigma}+\hc)+\omega_0\sum_ib_i^{\dagger}b_i \\
&&+g\sum_i(b_i+b_i^{\dagger})\Big(\sum_{\sigma}\hat
n_{i,\sigma}-1\Big).
\nonumber
\end{eqnarray}
$\elcre i{\sigma}$ creates an electron at lattice site $i$ with spin
$\sigma$,
and $b_i^{\dagger}$ a phonon with oscillator frequency $\omega_0$,
$\hat n_{i,\sigma}=\elcre i{\sigma}\elann 
i{\sigma}$. The electronic density is coupled to an optical phonon mode
with
coupling constant $g$. We have set the ionic mass to $M=1$ in
(\ref{holham}). The local 
oscillator displacement is related to the bosonic operators by  $\hat
x_i=(b_i+b_i^{\dagger})/\sqrt{2\omega_0}$, where $\hbar=1$.

For the DMFT at $T=0$ we solve the effective impurity
problem with the numerical renormalization 
group \cite{Wil75,BCP08} (NRG) adapted to the case with symmetry
breaking.\cite{BH09,BHD09}
For the logarithmic discretization parameter is $\Lambda=1.8$, and we keep about
1000 states at each iteration. The initial bosonic Hilbert 
space is restricted to a maximum of 50 states. We use a semi-elliptic density of
states (DOS) for the electrons $\rho_0(\epsilon)=\sqrt{4t^2-\epsilon^2}/(2\pi t^2)$
with bandwidth $W=4t$. In terms of the Hilbert transform $ {\rm
HT}[\rho_0](z)$ we have for the diagonal Green's function
\begin{equation}
G_{11}(i\omega_n)
= A_G {\rm HT}[\rho_0](\epsilon_+)+B_G {\rm HT}[\rho_0](\epsilon_-),
\label{G11}
\end{equation}
and for the off-diagonal part
\begin{equation}
G_{21}(i\omega_n)
= A_F {\rm HT}[\rho_0](\epsilon_+)+B_F {\rm HT}[\rho_0](\epsilon_-).
\label{G21}
\end{equation}
We have defined
$A_G=(\zeta_2(i\omega_n)+\epsilon_+(i\omega_n))/(\epsilon_+(i\omega_n)-\epsilon_-(i\omega_n))$,
$B_G=(\zeta_2(i\omega_n)+\epsilon_-(i\omega_n))/(\epsilon_-(i\omega_n)-\epsilon_+(i\omega_n))$
$A_F=\Sigma_{21}(i\omega_n)/(\epsilon_+(i\omega_n)-\epsilon_-(i\omega_n))$, and
$B_F=\Sigma_{21}(i\omega_n)/(\epsilon_-(i\omega_n)-\epsilon_+(i\omega_n))$, where
\begin{eqnarray}
\epsilon_{\pm}&=&\frac{\zeta_1(i\omega_n)-\zeta_2(i\omega_n)}{2}\pm \\
&&\frac12\sqrt{(\zeta_1(i\omega_n)+\zeta_2(i\omega_n))^2-4\Sigma_{21}(i\omega_n)  \Sigma_{12}(i\omega_n)},\nonumber
\end{eqnarray}
with $\zeta_{1}(z)=z+\mu-\Sigma_{11}(z)$ and
$\zeta_{2}(z)=z-\mu-\Sigma_{22}(z)$.  For the Nambu Green's functions
we have $G_{12}(i\omega_n)=G_{21}(i\omega_n)$ and
$G_{22}(i\omega_n)=-G_{11}(-i\omega_n)$. This implies
 $\Sigma_{12}(i\omega_n)=\Sigma_{21}(i\omega_n)$ and
$\Sigma_{22}(i\omega_n)=-\Sigma_{11}(-i\omega_n)$ for the self-energies. 
At half filling  $G_{11}(i\omega_n)$ and
$\Sigma_{11}(i\omega_n)$ are imaginary functions, whereas $G_{21}(i\omega_n)$ and
$\Sigma_{21}(i\omega_n)$ and $D(i\omega_m)$ and $\Sigma_{{\rm ph}}(i\omega_m)$
are real functions.  For the semi-elliptic DOS the Hilbert transform is given by
\begin{equation}
  {\rm
    HT}[\rho_0](z)=\integral{\epsilon}{-D}{D}\frac{\rho_0(\epsilon)}{z-\epsilon}=\frac{1}{2
    t^2}\Big(z-{\rm sgn}(\Imag(z))\sqrt{z^2-4t^2}\Big),
\end{equation}
where the square root of a complex number $w$ is given by
$\sqrt{r}\e^{i\varphi/2}$, where $\varphi=[0,2\pi)$, such that the imaginary
part of $\sqrt{w}$ is positive.

At finite temperature, we use
the continuous-time quantum Monte Carlo (QMC) method developed for
electron-phonon systems.\cite{AL07} To calculate $T_c$ we study the
susceptibility in the pairing channel $\chi(\vq,i\omega_n)$  in the limit
$\vq\to0$ and $i\omega_n\to 0$. It can be expressed in terms of the irreducible vertex in the particle-particle channel $\Gamma^{(\rm
  pp)}$, which is calculated in the QMC procedure.\cite{GKKR96}  It is then sufficient to analyze when the largest
eigenvalue of the symmetric matrix,
\begin{equation}
M_{n_1,n_2}= \frac{1}{\beta}\sqrt{\tilde\chi^0(i\omega_{n_1})}[\Gamma^{(\rm
    pp)}(i\omega_{n_1},i\omega_{n_2};0)]\sqrt{\tilde\chi^0(i\omega_{n_2})},
\label{symmatrixinst}
\end{equation}
exceeds one. We have defined the pair propagator,\cite{GKKR96}
\begin{equation}
  \tilde\chi^0(i\omega_{n_1})=\frac{G(i\omega_{n_1})-G(-i\omega_{n_1})}{\zeta(-i\omega_{n_1})-\zeta(i\omega_{n_1})},
\label{tchi0gen}
\end{equation}
$\zeta(i\omega_n)=i\omega_n+\mu-\Sigma(i\omega_n)$,
where $G(i\omega_{n_1})$ is the local lattice Green's function in the normal state.

\subsection{ME approach}

\begin{figure}[!t]
\centering
\includegraphics[width=0.22\textwidth]{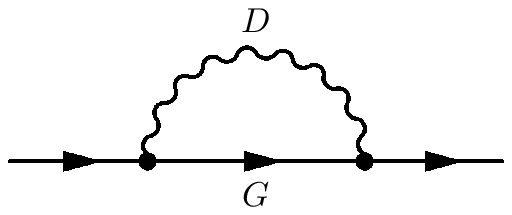}
\hspace{0.2cm}
\includegraphics[width=0.2\textwidth]{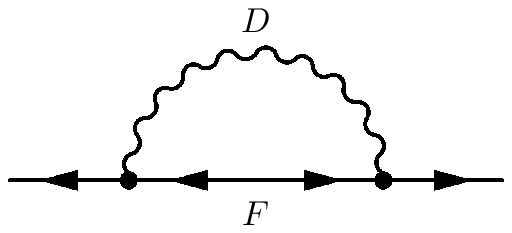}
\caption{Diagrams for diagonal and offdiagonal self-energy in ME theory
  involving the full electronic Green's function $G$ and anomalous part $F$
  and the full phonon propagator $D$.}        
\label{fig:MEdiagrams}
\end{figure}

The matrix equation for the electronic self-energies in ME theory (see
Fig. \ref{fig:MEdiagrams}) reads
\begin{equation}
\underline{\Sigma}(i\omega_n)=
-\frac{1}{\beta}\sum_{m}g^2 \tau_3
\underline{G}(i\omega_m+i\omega_n)\tau_3 D(i\omega_m) ,
\label{MEeq}
\end{equation}
where
\begin{equation}
\underline{G}_{\vk}(i\omega_n)^{-1}=\underline{G}^0_{\vk}(i\omega_n)^{-1}-
\underline{\Sigma}(i\omega_n)
\end{equation}
with $G_{ij}(i\omega_n)=\sum_{\vk}G_{ij,\vk}(i\omega_n)$ and
$\underline{G}^0_{\vk}(i\omega_n)^{-1}=i\omega_n
-\tau_3(\epsilon_{\vk}-\mu)$. Notice that we use the form valid for
large coordination number, such that only local lattice Green's functions
enter. These are calculated with the same semi-elliptic DOS as in Eqs. 
(\ref{G11}) and (\ref{G21}) for the DMFT calculation. Thus by ME theory we
mean the diagrammatic theory, which neglects all vertex corrections to
(\ref{MEeq}), and this is compared to the full DMFT results. No further
approximation such as assuming a constant density of states or large bandwidth
are made. The latter are common approximations in the literature on ME theory,\cite{Car90,MC08}
whose accuracy can be analyzed in an expansion in $\omega_{\rm
  ph}/W$. This is, however, not the subject of the present paper, where we focus entirely on the accuracy of 
the theory when vertex corrections are neglected. 

The pairing function reads $\alpha^2F(\omega)=\rho_0g^2\rho^D(\omega)$
for the
Holstein model. In the non-interacting limit we have
$\rho^D_0(\omega)=\delta(\omega-\omega_0)-\delta(\omega+\omega_0)$, 
which in Eq. (\ref{eq:deflambda}) gives $\lambda_0=\rho_02g^2/\omega_0$
purely in terms of bare parameters. This quantity was used in model studies
and denoted by $\lambda$.\cite{differentlambda} However, $\lambda$ as
defined in Eq. (\ref{eq:deflambda}) is given for the {\em interacting}
system.\cite{MK82,Mar90,DAM08}  
Then the phonons are renormalized  via the self-energy
$\Sigma_{{\rm ph}}(i\omega_m)$,
\begin{equation}
 D(i\omega_m)^{-1}=D^0(i\omega_m)^{-1}-\Sigma_{{\rm ph}}(i\omega_m),
\end{equation}
where $D^0(i\omega_m)=2\omega_0/[(i\omega_m)^2-\omega_0^2]$.
The lowest order contribution to the phonon self-energy  
is
\begin{equation}
  \Sigma_{{\rm
      ph}}(i\omega_m)=\frac{2g^2}{\beta}\sum_{n}G(i\omega_n)G(i\omega_m+i\omega_n).
\label{phononse}
\end{equation}
As in Ref. \onlinecite{CC03} we will call Eqs. (\ref{phononse}) and
(\ref{MEeq}) {\it self-consistent} ME approximation.
In the superconducting state an additional contribution from the
off-diagonal
Green's function could be taken into account, which is however small and
it will be neglected in the following. 

We define the peak of the interacting phonon spectral function
$\rho^D(\omega)$ as the effective phonon scale $\omega_{\rm ph}=\omega_0^r$.
There is then a mapping of the bare dimensionless parameters
$\lambda_0,\omega_0/t$ to the effective parameters
$\lambda,\omega_0^r/t$. $\lambda$ exceeds the bare 
$\lambda_0$ due
to the phonon renormalization, $\omega_0\to\omega_0^r$, and due to
the increased lattice fluctuations as shown in the 
identity valid at $T=0$,
\begin{equation}
w_D=\integral{\omega}{0}{\infty} \rho^D(\omega)=2\omega_0\expval{\hat
x^2}{},
\label{eq:defwD}
\end{equation}
which is generally larger than one. For a sharply peaked phonon spectrum
the first moment sum rule,
\begin{equation}
 \integral{\omega}{-\infty}{\infty} \omega \rho_{D}(\omega)=2\omega_0 ,
\end{equation}
implies $w_D\simeq \omega_0/\omega_0^r$. From an estimate for the
phonon softening due to the lowest order diagram,
$\omega^r_0/\omega_0=\sqrt{1-a\lambda_0}$, and Eq. (\ref{eq:deflambda}) one can then obtain the result   
$\lambda=\lambda_0/(1-a\lambda_0)$.\cite{Mig58,MK82,DAM08} In three
dimensions $a=2$, and for a semi-elliptic DOS in the limit of large
dimensions we have $a=8/3$.

In order to calculate the gap at $T=0$ we solve equation (\ref{MEeq}) both 
by introducing spectral functions and analytic continuation to the real axis
and for comparison directly on the imaginary axis. 
$D(i\omega_m)$ can be calculated self-consistently via Eq. (\ref{phononse})
or taken as an input from DMFT calculations. The latter type of calculation is
termed ME+ph. In the ME theory $T_c$  is calculated by first finding the local
lattice Green's function $G(i\omega_n)$ in the normal phase using
\begin{equation}
    \Sigma(i\omega_n)=-\frac{g^2}{\beta}\sum_{m}G(i\omega_m+i\omega_n)D(i\omega_m),
\end{equation}
and then employing
Eq. (\ref{symmatrixinst}) with 
\begin{equation}
  \Gamma^{(\rm
    pp)}(i\omega_{n_1},i\omega_{n_2};0)=-g^2D(i\omega_{n_1}-i\omega_{n_2}).
\label{elphvert}
\end{equation}

\section{Results}
\begin{figure}[!thbp]
\centering
\subfigure[]{\includegraphics[width=0.39\textwidth]{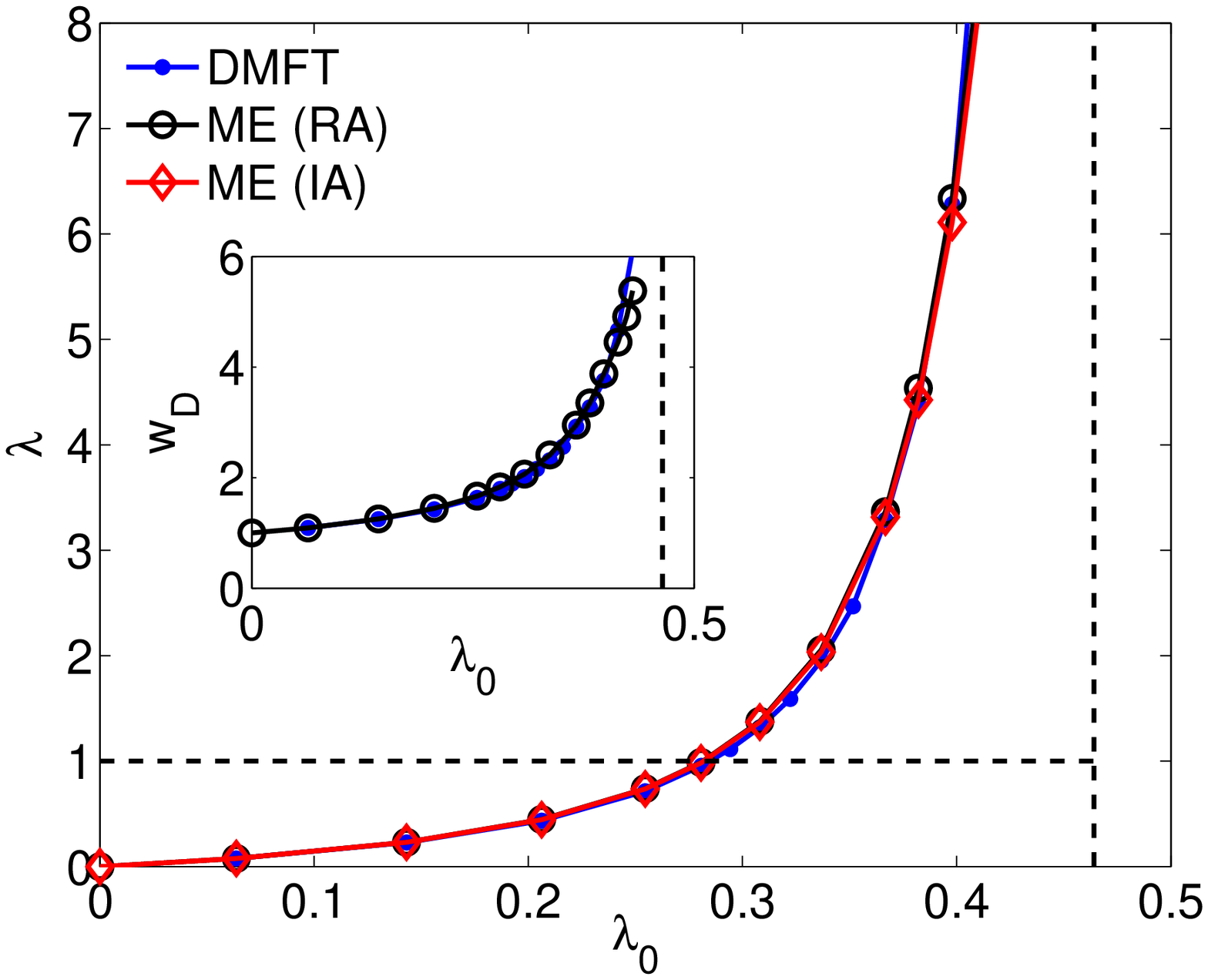}}
\subfigure[]{\includegraphics[width=0.42\textwidth]{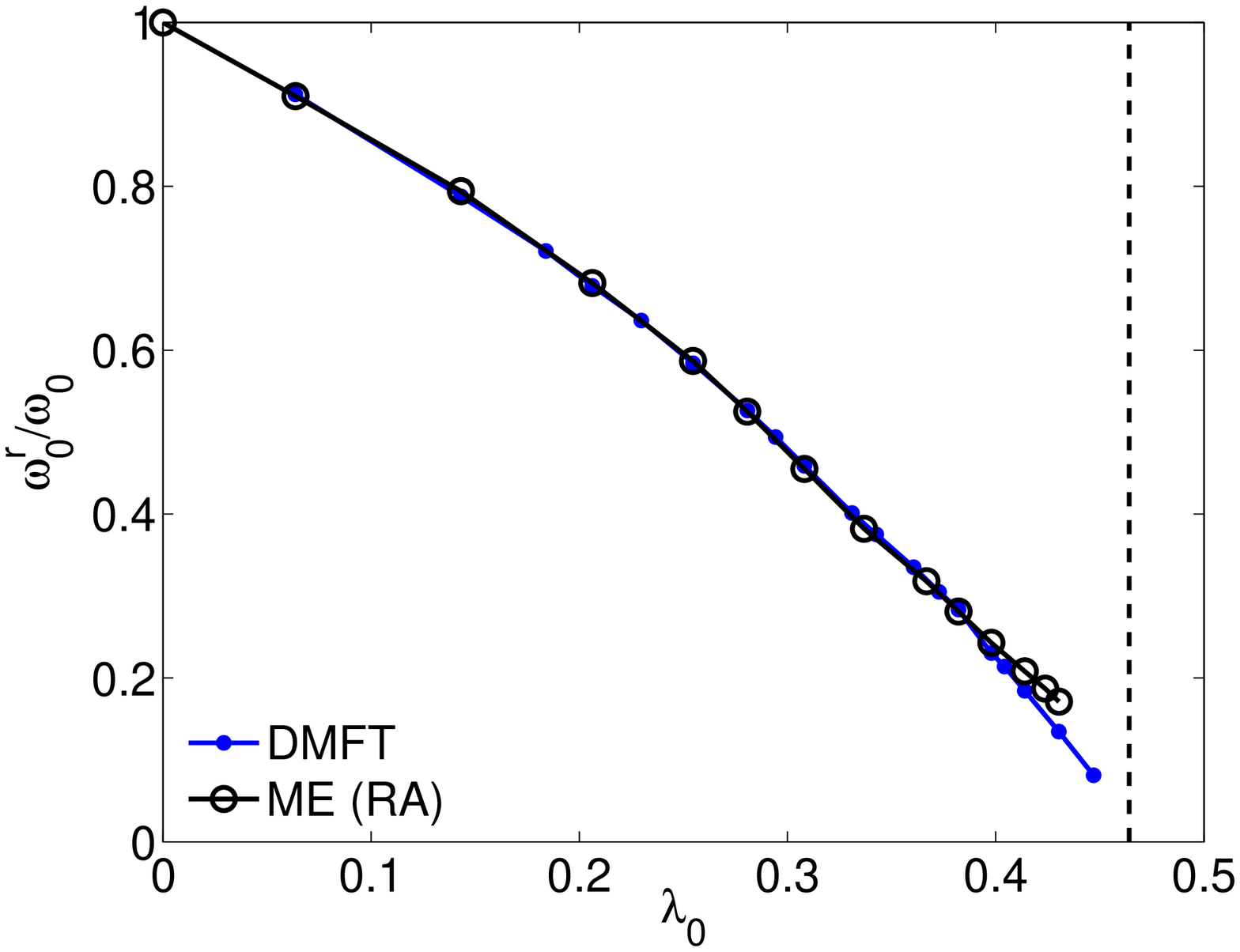}}
\vspace*{-0.5cm}
\caption{(Color online) Comparison of the selfconsistent ME and DMFT
results for renormalized
 quantities as a function of the bare $\lambda_0$: (a) The
 effective coupling $\lambda$ as defined by Eq. (\ref{eq:deflambda}), 
 inset $w_D$ as given by Eq. (\ref{eq:defwD}). (b) The ratio of
renormalized
 and bare phonon frequency $\omega_0^r/\omega_0$.}       
\label{fig:lambdar}
\end{figure}

Let us first of all establish how the bare and effective quantities are
related at $T=0$. 
At half filling for fixed $\omega_0=0.1 t$, we plot $\lambda$ in
Fig. \ref{fig:lambdar} (a)
and $\omega_0^r/\omega_0$ in Fig. \ref{fig:lambdar} (b) both as function
of $\lambda_0$.
We show the results from self-consistent ME theory on the real
axis (RA) and on the imaginary axis (IA) 
in comparison with the full DMFT-NRG result. 

\noindent
$\lambda$ increases slowly for $\lambda_0\le 0.3$ up to
values around one. Then it rises more rapidly close to
values of $\lambda_0$ where in the normal state a metal to bipolaronic (BP) insulator
transition had been found at $\lambda_0^c\simeq 0.464$ (shown as a vertical
line).\cite{BZ98,MHB02} The behavior is qualitatively similar to the analytic
estimate above, however, as $\lambda_0^c$>$1/a$ the latter 
is a substantial overestimate and 
diverges too quickly. The region of most interest for our   
purpose is $\lambda\sim 1-3$, typical values for strong coupling
superconductors. This corresponds to $\lambda_0\sim 0.3 -0.37$ in terms of
bare parameters.  The values for $\lambda$ obtained in the self-consistent ME
theory compare well to the DMFT results for smaller values of $\lambda_0\le
0.3$, and then start to overestimate this quantity slightly. For 
values of $\lambda_0$ closer to the BP transition self-consistent ME
underestimates $\lambda$. We
also compare the effective phonon frequency which decreases with
$\lambda_0$ towards zero when $\lambda_0$ approaches $\lambda_0^c$. This
quantity compares well to the DMFT result for a considerable range of 
$\lambda_0$, but starts to deviate for $\lambda_0>0.38$ or $\lambda\gtrsim
4$.

\begin{figure}[!t]
\centering
\subfigure[]{\includegraphics[width=0.42\textwidth]{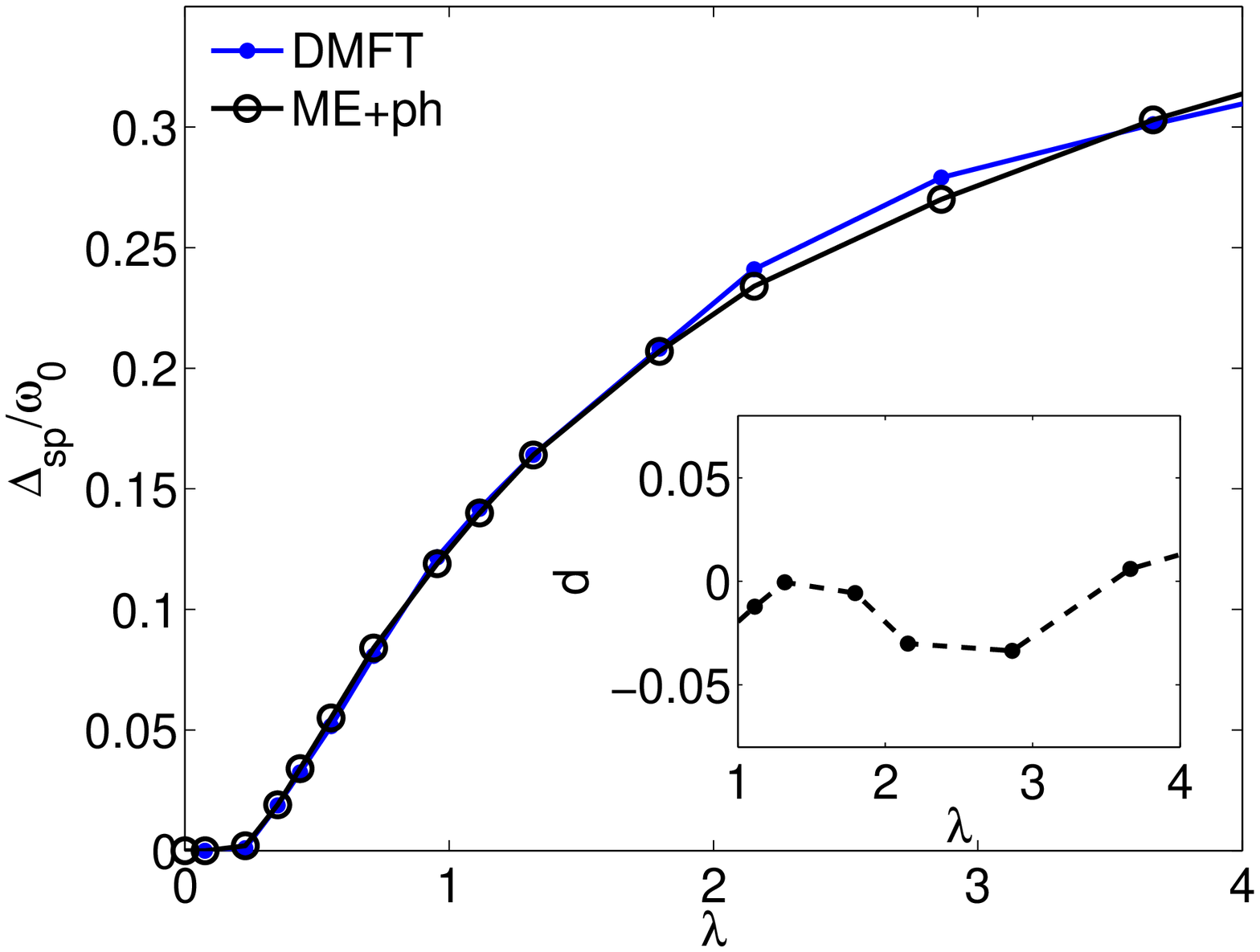}}
\subfigure[]{\includegraphics[width=0.42\textwidth]{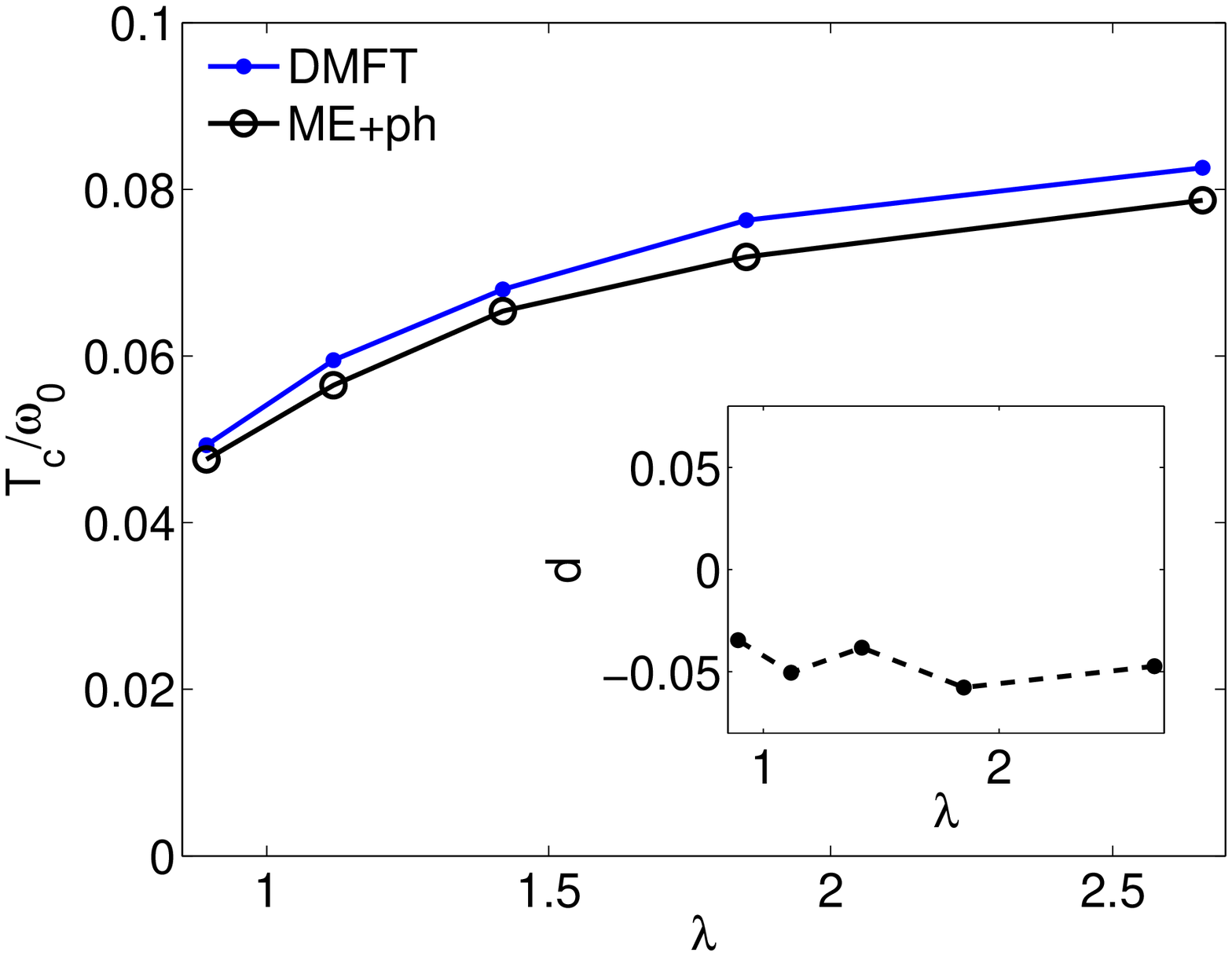}}
\vspace*{-0.5cm}
\caption{(Color online) Comparison of quantities for the superconducting
state  as a function of $\lambda$ obtained from DMFT and ME calculations with phonon input from
 DMFT (ME+ph): (a) The spectral gap $\Delta_{\rm
   sp}$; the inset shows the relative deviation $d$. (b) The
 critical temperature for the onset of superconductivity $T_c$.}
\label{fig:reslambdadep}
\end{figure}
\noindent

We can also calculate electronic properties like the quasiparticle
weight $z$
or the offdiagonal self-energy $\Sigma^{\rm off}(0)$ which roughly
determines the
spectral gap at zero temperature, $\Delta_{\rm sp}\simeq z \Sigma^{\rm
 off}(0)$. Then one finds good agreement for small coupling and
moderate deviations between DMFT and  self-consistent ME theory 
in the intermediate coupling regime, and close to the bipolaronic
transition, similar to the results for $z$ and $\omega_0^r$ which have
been obtained by Ciuchi and Capone \cite{CC03} in the normal state.

Our main objective is to test the validity of the ME theory at strong
coupling. Hence, we compare the results for the superconducting properties
$\Delta_{\rm sp}$  and $T_c$ obtained from the ME+ph calculations with the full DMFT results. 
In Fig. \ref{fig:reslambdadep} (a) we show $\Delta_{\rm sp}$ as extracted from
the spectral function computed from ME+ph calculations on the real axis
and the corresponding DMFT result. Notice that 
the results are plotted as a function of $\lambda$ now.

\begin{figure}[!thbp]
\centering
\includegraphics[width=0.42\textwidth]{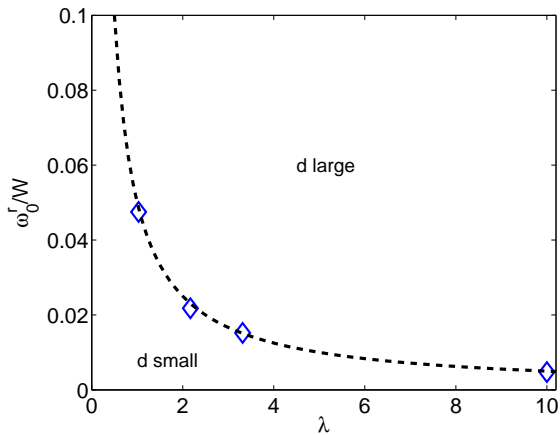}
\vspace*{-0.3cm}
\caption{(Color online) Guideline for the quantitative reliability of ME theory. The
 diagram shows points (diamonds) in the $\omega_0^r/W$-$\lambda$-plane where the
 deviation $d$ between the DMFT and ME+ph theory is $d\simeq 0.1$. The  dashed
 line is given by the functional form 
$\omega_0^r/W=c/\lambda$, which follows from an analytical estimate where vertex
correction exceed a certain value. $c=0.05$ was used.}       
\label{fig:reliability}
\end{figure}
\noindent

We find very good agreement for small values of $\lambda <1$, then a
regime where ME+ph slightly underestimates the value for the gap, before it 
exceeds the DMFT result for larger values of $\lambda$. By inspecting
the relative deviation $d=(\Delta_{\rm sp}^{\rm  ME}-\Delta_{\rm sp}^{\rm 
 DMFT})/\Delta_{\rm sp}^{\rm ME}$ plotted as an inset we see that there
is an agreement in the regime $\lambda\sim 1-3$ better than 10\%. 
At very large values of $\lambda$, $\Delta_{\rm sp}$ from ME+ph
increases stronger than the DMFT result. 
For similar parameters we have also calculated the critical temperature
$T_c$ as deduced from the Bethe-Salpeter equation of the uniform pair 
susceptibility.\cite{HGC03} The comparison of ME
theory and DMFT-QMC result is shown in Fig.~\ref{fig:reslambdadep}~(b).
Good agreement is found in the relevant range for $\lambda$. 
DMFT-QMC systematically slightly underestimates the phonon
renormalization, which accounts partly for the discrepancy of $T_c$ from the ME+ph calculations.  
In a related approach Marsiglio found for a 4$\times$4 cluster
that the self-consistent ME theory agrees well with QMC calculations for
the pairing susceptibility.\cite{Mar90}

By doing similar comparisons for different bare parameters we mapped
out for which values of the effective parameters $\lambda$ and $\omega_0^r/W$
DMFT and ME+ph show good agreement, i.e. $d\lesssim 0.1$. 
The results are shown in Fig.~\ref{fig:reliability} and can be well understood
in terms of the effective expansion parameter of ME theory
$\lambda\omega_0^r/W$, which should not exceed 0.05 for good accuracy. The
results can serve as a guideline for the application of ME theory with
reliable phonon input. 

\section{Conclusions}
We have assessed the validity of the ME theory.
We calculated accurately the effective coupling strength $\lambda$
in terms of the bare coupling strength $\lambda_0$. For intermediate $\lambda_0$ the 
system is close to a bipolaronic metal-insulator transition and
$\lambda$ is very strongly enhanced. Close to this point ME theory breaks down. 
However, for typical values for strong coupling superconductors, $\lambda\sim 1-3$,
the ME theory is very accurate for small values of $\omega_{\rm ph}/W$.
This result is demonstrated explicitly for the Holstein model in the limit of large
dimensions, where most of the spectral weight of the pairing function is
located at $\omega_{\rm ph}$. 
We expect that this result is also applicable for more general forms of
pairing functions $\alpha^2F(\omega)$ in three dimensions with an appropriate
cut-off scale $\omega_{\rm ph}$.
In many applications of ME theory a momentum average over the Fermi surface is taken,
such that the situation is similar to the one studied here. However, the
momentum dependence can be important in certain cases especially for lower dimensional
materials. For instance in Ref. \onlinecite{GPS95}, the momentum dependence of
vertex corrections and their effect on $T_c$ was analyzed.

\subsection*{Acknowledgments}
We wish to thank F. F. Assaad, A.C. Hewson, G.
Sangiovanni, and R. Zeyher for helpful discussions, and to M. Kulic for
pointing out Ref. \onlinecite{MK82} to us. JH acknowledges support from the grant
NSF DMR-0907150.

\bibliography{artikel,biblio1,footnote}

\begin{thebibliography}{25}
\bibitem[{dif()}]{differentlambda}
\bibinfo{note}{The definition of $\lambda$ can vary from ours by numerical
  prefactors, for instance $\pi$ in Ref. \onlinecite{BZ98} or $\pi/2$ in Ref.
  \onlinecite{CC03}. The definition in Ref. \onlinecite{Ale01} corresponds to
  ours. As emphasized in this paper bare and renormalized quantities have to be
  distinguished.}
\end{thebibliography}

\end{document}